\documentclass[conference]{IEEEtran}
\IEEEoverridecommandlockouts
\usepackage{cite}
\usepackage{verbatim}
\usepackage{braket}
\usepackage{amsmath,amssymb,amsfonts}
\usepackage{algorithmic}
\usepackage{graphicx}
\usepackage{textcomp}
\usepackage{xcolor}
\usepackage{listings}
\usepackage[hidelinks]{hyperref}
\def\BibTeX{{\rm B\kern-.05em{\sc i\kern-.025em b}\kern-.08em
    T\kern-.1667em\lower.7ex\hbox{E}\kern-.125emX}}
\begin{document}

\title{Entangling Disciplines: Causality, Entropy and Time-Travel Paradoxes on a Quantum Computer\\
\thanks{I thank the Heilbronn Institute for Mathematical Research and IBM Quantum for their support.}
}

\author{\IEEEauthorblockN{Maria Violaris}
\IEEEauthorblockA{\textit{University of Oxford} \\
Oxford, United Kingdom \\
maria@violaris.com}
}

\IEEEoverridecommandlockouts \IEEEpubid{\begin{minipage}{\textwidth}\ \\[12pt]\ \copyright2024 IEEE. Personal use of this material is permitted. Permission from IEEE must be obtained for all other uses, in any current or future media, including reprinting/republishing this material for advertising or promotional purposes, creating new collective works, for resale or redistribution to servers or lists, or reuse of any copyrighted component of this work in other works.\hfill \end{minipage}}

\maketitle

\begin{abstract}
Merging disciplines has led to incredible learnings and breakthroughs throughout history, including the discovery of quantum computing: a cross between computation and quantum physics. In this paper, I will discuss how we can cross quantum computing with topics in fundamental physics. This leads to fruitful, interactive learning opportunities that fuse deep open physics problems with key insights about quantum information science. By outlining quantum circuit experiments that can be run on current and near-term quantum computers, I demonstrate how to help learners engage with principles in special relativity, general relativity and thermodynamics. In turn, these connections can advance their understanding of quantum computing. Learners can further explore the quantum computing activities in this paper via the Quantum Paradoxes content series of videos, blogs and code tutorials that I created with IBM Quantum.
\end{abstract}


\section{Introduction}

Quantum computing is an inherently interdisciplinary field, both in terms of the disciplines needed to build quantum computers (e.g. physics, maths, computer science, engineering), and those where these computers have potential applications (e.g. chemistry, materials science, finance). Here we will be focusing on a different kind of interdisciplinarity: where quantum computing meets fundamental principles in fields of physics beyond quantum mechanics. 

Here I propose quantum computing-based activities for learners to test fundamental physics principles. Each activity gives learners insights into the constraints and scope of quantum computers, in relation to special relativity, general relativity, and thermodynamics. They will discover the implications of quantum computing for broader aspects of physics such as entropy, spacetime, and how modified physical laws could enable more powerful forms of computation. The three main activities I present draw upon the ``Quantum Paradoxes" content series that I created with IBM Quantum, which has a YouTube video, blog and Qiskit code tutorial for each instalment \cite{violaris2023quantum, Qiskit}. Alongside each activity, this paper includes suggestions for further extensions to explore the crossover of quantum computing with these topics. 

These activities have been used for lectures and workshops with High School students, teachers, undergraduates and graduates. Learners are expected to know the basics of quantum circuits and quantum computing. The physics and maths explanations can be adapted for different backgrounds, making use of the visual, interactive quantum circuit implementations as a pedagogical tool. 

First I present a quantum circuit implementation of the EPR paradox thought experiment. I explain how we can use the results of running the circuit to demonstrate core principles of special relativity, no faster-than-light communication and no-signalling, ensuring the consistency of causality. Then I show how to use the circuit to make stronger conclusions regarding the principle of locality, and its consistency with general relativity. 

\IEEEpubidadjcol

Then I consider a quantum-circuit simulation of a quantum Maxwell's Demon. The classical Maxwell's Demon thought experiment demonstrates how a system (hypothetical demon) with enough microscopic control could use a blank memory as a resource for extracting useful work from random motion. The extension of the thought experiment to a quantum setting shows how classical information theory formulations of Shannon entropy translate to von Neumann entropy in quantum information theory. This demonstrates how key aspects of thermodynamics translate to a quantum, microscopic, unitary regime; the significance of ancillary systems (such as pure qubits) for enabling certain useful quantum transformations; and constraints on the resources required for erasure of quantum information in quantum computers. 

Finally I show how to use quantum computers to simulate time-loops, otherwise known as closed-timelike-curves (CTCs). CTCs are permitted by the theory of general relativity, however they lead to logical paradoxes. One of these is the Grandfather Paradox, in which someone can use a CTC to go back in time and kill their own grandfather before they were born. Then, they were never alive to kill their grandfather in the first place, causing a contradiction. Using quantum theory, the paradox can be resolved \cite{deutsch1991quantum}. With the quantum circuit simulation, learners can test for themselves the consistency of CTCs with a version of the Grandfather Paradox. 

Furthermore, by seeing how parts of the simulation require classical control to artificially insert non-linearity into the system, learners can test directly how CTCs would give us access to computational power beyond that allowed by standard quantum theory \cite{aaronson2009closed}. They can test this by using the simulation to distinguish non-orthogonal input states with a single measurement \cite{brun2009localized}. In particular, they can construct a circuit to distinguish between four non-orthogonal input states, simulating how CTCs would break the security of the BB84 cryptographic protocol \cite{brun2009localized}. 

Additionally, the quantum circuit demonstrates open problems with the fundamental physics of CTCs. One of these is how quantum CTCs violate the principle of locality introduced by the first activity about the EPR paradox. Another is the ``knowledge paradox", where information can appear in a CTC without explanation \cite{deutsch1991quantum}. 

Exploring the intersections of quantum computing with other disciplines can help quantum computing learners gain novel insights into other fields of physics, such as how to extend thermodynamics to a quantum setting. Simultaneously, learners can gain inspiration from other fields to clarify their understanding of the scope, power and potential of quantum computing. For example, by understanding energetic limits of information processing; the underlying dynamics of information flow via entanglement; and constraints on computational power imposed by CTCs. 

\IEEEpubidadjcol

\section{Bell's theorem, causality and locality}

In 1935, Albert Einstein, Boris Podolsky and Nathan Rosen proposed a thought experiment that came to be known as the EPR paradox. By analysing the properties of two entangled systems, they concluded that quantum mechanics is an incomplete theory, as otherwise the implications of entanglement seemed to conflict with core principles of Einstein's relativity theories. Later, John Bell proposed a novel approach to analysing measurements of quantum systems to see if the outcomes violate statistical inequalities. If these inequalities are violated, it rules out a class of theories known as ``local hidden variable models", which are potential candidates for modifying or completing quantum theory. Experimental implementations to test out Bell inequalities, and indeed rule out the impossibility of local hidden variable models describing reality, were then carried out by John Clauser, Alain Aspect and Anton Zeilinger \cite{brassard2023profile}. These had such fundamental significance for physics that they won the 2022 Nobel Prize. To understand the nature of entanglement, we need to consider also the limitations of Bell's theorem, and the local formulations of quantum theory that are not ruled out by violations of Bell inequalities.

A common popular idea is that given two entangled quantum systems, quantum theory implies that measuring one of the systems has an instantaneous effect on the other system, however far apart they are. Often attributed to Einstein's phrase ``spooky action at a distance", such behaviour would violate Eintein's \textit{principle of locality}. Einstein's principle of locality says that, for any two systems A and B, ``the real factual situation of system A is independent of what is done with the system B, which is spatially separated from the former" (paraphrased from \cite{schilpp1959albert}). There are several important features of physics that follow from this principle: \\

1) No-signalling and no faster-than-light communication: If two systems are separated by space, and nothing you do to one can affect the spatially separated one, then there is no way to communicate a message from one to the other (``signal") without something physically moving between them. Combined with the principle in special relativity that no physical entity can travel faster than light, no-signalling implies the impossibility of faster-than-light communication. \\
2) No instantaneous influences across space: instantaneous effects does not just rule out physically observable ones that could be used to communicate, but rather \textit{any influence at all} on the ``real, factual situation" of a system.  \\
3) Local, complete descriptions: For the real, factual state of a system to be independent from anything spatially separated from it, there must exist a complete mathematical description for the properties of an individual physical system. 

At first glance, Einstein's principle of locality seems to be contradicted by many expositions of the nature of entanglement. Common phrases are along the lines of: ``if two quantum systems are entangled, then measuring one system can instantaneously affect the other system, even if it is on the opposite side of the universe". Using the quantum circuit implementation of measurements on an entangled pair of qubits, I will explain how we can explicitly see the manifestation of (1) no-signalling; (2) no instantaneous influences; and (3) local, complete descriptions of entangled qubits. This is all done without reference to the local hidden variable models that are ruled out by Bell's theorem and violations of Bell inequalities. 

\subsection{Quantum circuit activity}\label{epr-circuit}

\begin{figure}[htbp!]
\centerline{\includegraphics[width=0.4\textwidth]{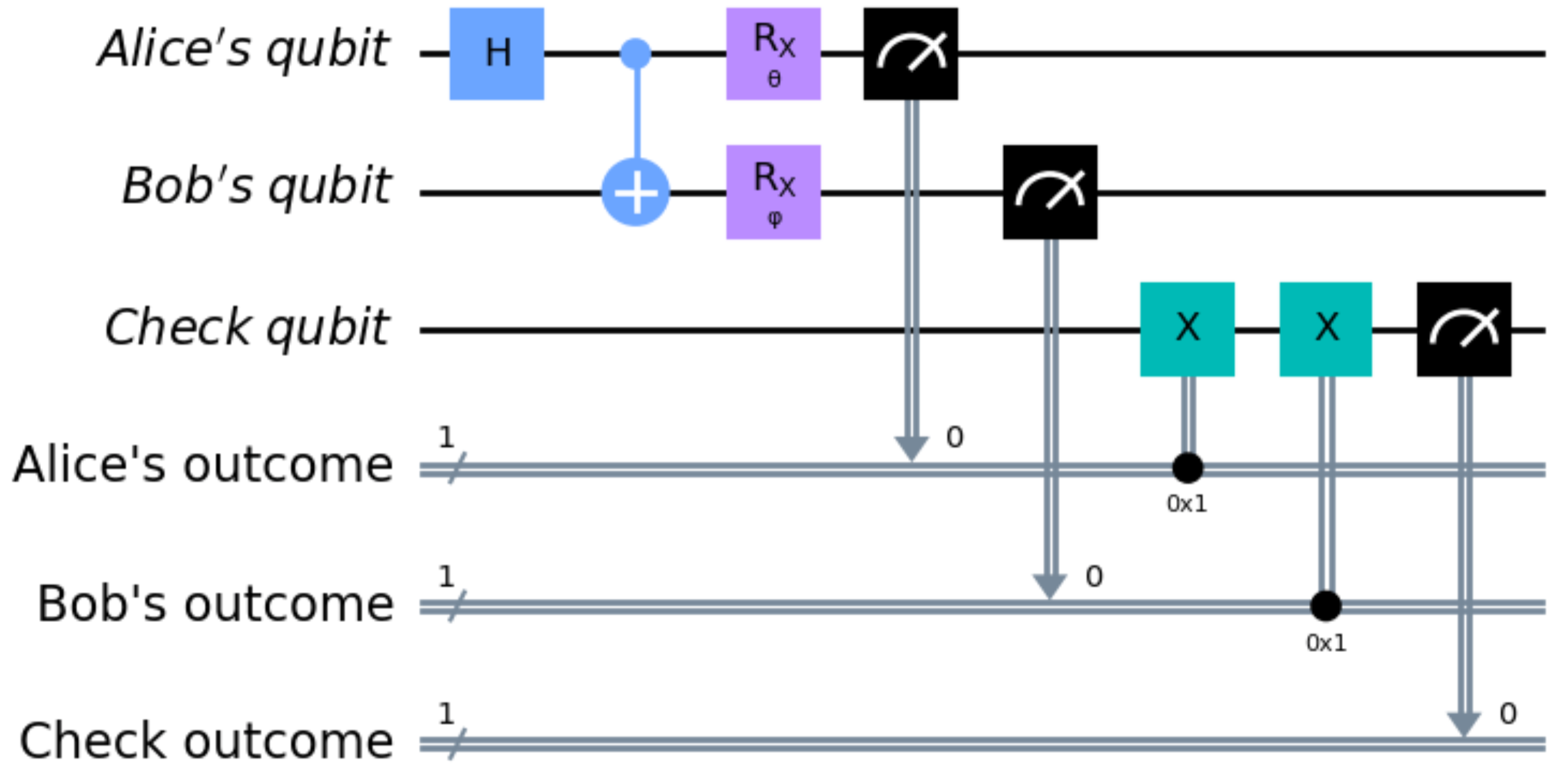}}
    \caption{Quantum circuit for measuring correlations of an entangled pair of qubits. Alice and Bob share an entangled pair; they each apply an arbitrary single-qubit X-rotation, paramaterised by $\theta$ and $\phi$ respectively; and they compare whether their outcomes are the same or different via a "parity check" using the check qubit.}
    \label{fig:epr-classical}
\end{figure}

A quantum circuit for a version of the EPR-thought experiment, where entangled particles are separated and measurements can be done on each of them, is shown in Figure \ref{fig:epr-classical}.  This and the other circuit figures in this paper were created using Qiskit \cite{Qiskit}.

Here we imagine that two experimenters, Alice and Bob, share an entangled pair of qubits. These are the first two qubits in the circuit. Then they can each choose a basis in which to measure their qubit, which I have captured by adding a paramaterised single-qubit gate before each of their Z-measurements. For simplicity, I have made both these gates be X-rotation gates, with Alice's characterised by $\theta$ and Bob's by $\phi$. In general they could both be arbitrary unitary gates U, where each U is defined by three parameters, and the same explanation will hold. 

I have then introduced a ``check qubit", which stores the outcome of a parity check on Alice and Bob's measurement outcomes, implemented by a pair of CNOT gates controlled on Alice and Bob's measurement outcomes and targeted on the check qubit. If Alice and Bob's measurement outcomes are the same, the check qubit will stay in the $\ket{0}$ state, while if Alice and Bob's measurement outcomes are different, it will change to the $\ket{1}$ state.

\subsubsection{\textbf{Quantum and classical correlations}}

The initial Hadamard and CNOT gates prepare Alice and Bob's entangled pair in the maximally entangled Bell state $\frac{1}{\sqrt{2}}(\ket{00}+\ket{11})$. It is clear that if they both do Z-measurements on their qubits, then they will always get fully correlated outcomes of either 0,0 or 1,1. The level of correlation between the qubits demonstrated by such a measurement can also be found in classical systems: for example, if Alice and Bob were sharing a pair of socks that were either pink or blue, then their measurements of their socks will always either be pink, pink or blue, blue. 

However, entangled qubits are correlated in a stronger way than classical systems such as a pair of socks. This is manifested in the following behaviour of the quantum circuit: whatever basis Alice and Bob choose to measure their entangled pair of qubits in, then as long as they choose the same basis, their measurement outcomes will be fully correlated.  For example, if they both measure in the X-basis by applying Hadamard gates and then standard Z-measurements, they will again get outcomes of 0,0 or 1,1, corresponding to measurements of the qubits in states $\ket{+}$, $\ket{+}$ or $\ket{-}$,$\ket{-}$ respectively (i.e. fully correlated outcomes in the X-basis). 

Classical systems can also be correlated in multiple degrees of freedom. Continuing the sock comparison, imagine that the pink pair of socks is large, and the blue pair is small. Then, Alice and Bob will always measure the property of size on their individual socks as being fully correlated, in addition to the property of colour. However, there is a fundamental difference between the classical properties of the socks (colour and size), and the quantum properties of the entangled pair (Z-basis and X-basis measurement outcomes). Pieces of information about the colour and size of a sock can be retrieved simultaneously, whereas pieces of information about the outcome of measuring a qubit in the Z-basis and X-basis cannot be retrieved simultaneously. This is a manifestation of Heisenberg's uncertainty principle, which famously says that position and momentum cannot be measured simultaneously. Mathematically, the constraint applies for any pair of observables that do not commute, which is a feature unique to quantum systems. 

This property is what makes it strange for Alice and Bob's measurement outcomes to be fully correlated, in whichever basis they do their measurements. It seems like Alice's choice of measurement basis has some instantaneous effect on Bob's qubit, since Bob's qubit always gives the correct measurement outcome such that it is correlated with Alice's. This effect is manifest in the quantum circuit. Even though Alice and Bob are free to choose the basis in which they measure their own qubit independently, by varying the parameters $\theta$ and $\phi$, somehow their measurement outcomes conspire to be the same when the parameters coincide, such that the check qubit is always $\ket{0}$.

\subsubsection{\textbf{Bell non-locality}}

Bell's theorem is often quoted to support this viewpoint of there being an influence on Bob's qubit caused by Alice's choice of measurement basis, and/or vice-versa. Bell's theorem considers a class of local theories, called ``local hidden variable models". These assumptions place constraints on Alice and Bob's measurement outcomes, which are violated by quantum mechanics, showing that quantum mechanics cannot be a local hidden variable-model. The type of locality ruled out by ruling out this class of local theories is called Bell-non-locality. 

\subsubsection{\textbf{EPR thought experiment as a unitary quantum circuit}}

However, Bell's theorem only rules out one class of local models for quantum theory \cite{deutsch2000information}. To explain the alternative local formulation, we need to consider an alternative quantum circuit, shown in Figure \ref{fig:info-flow}. This is identical to the previous quantum circuit of Figure \ref{fig:epr-classical}, except we have changed the classical registers of Alice and Bob's measurement outcomes to quantum registers. Then Alice and Bob's measurement operations become CNOT gates, such that if the control is $\ket{0}$, the memory will stay in the $\ket{0}$ state, and if the control is $\ket{1}$, the memory will be flipped to the $\ket{1}$ state, such that each memory qubit stores the outcome of measuring the entangled qubit. 

Switching measurements for CNOT gates at the end of a circuit makes no difference to the final distribution of measurement outcomes, known as the ``principle of deferred measurement" in quantum computing. However, a key difference to the Bell-type approach is that here we do not assume that Alice and Bob have single, probabilistic measurement outcomes. Our overall description of events here is unitary, maintaining the overall entangled superposition of the multiple measurement outcomes that Alice and Bob could retrieve, freeing it from the assumptions of Bell's theorem. This is why the local account being explained here is fully consistent with measurement outcomes that violate Bell inequalities. 

\subsubsection{\textbf{Local and complete information from observables}}

Now to explicitly see the local account of this thought experiment, we need to shift from the usual description of quantum states using a global statevector to an explicitly local description of quantum states using observables. The global statevector description of states is called the \textit{Schrödinger picture}, while describing quantum states using observables is called the \textit{Heisenberg picture}. These two pictures are mathematically equivalent in the sense that they give the same predictions about the distributions of measurement outcomes in quantum mechanics, but only the Heisenberg picture makes the local formulation of quantum mechanics explicit. 

To understand the difference between the global statevector and observables, consider the expression for the expectation value of an observable $\hat{O}$ after applying some unitary gate $U$ to an initial state $\ket{0}$: $\bra{0}U^\dagger \hat{O} U \ket{0}$. $\hat{O}$ could be e.g. the Pauli $\hat{X}$ operator, if we want to find the average outcome of doing X-measurements, or the Pauli $\hat{Z}$ operator, if we want to find the average outcome of doing Z-measurements. In the Schrödinger picture, we would first find the global statevector $U\ket{0}$, and then sandwich the operator between the bra- and ket- to work out the expectation value: $(\bra{0}U^\dagger) \hat{O} (U \ket{0})$. In the Heisenberg picture, we first evolve the operator using the unitary $U$ and $U^\dagger$, and then sandwich this between the bra- and ket- of the initial statevector: $\bra{0}(U^\dagger \hat{O} U) \ket{0}$. The final expressions are the same. 

The properties of an individual quantum system can be fully described by the observables of that system. More specifically, we can form a vector of the $\hat{X}$, $\hat{Y}$ and $\hat{Z}$ observables of a system — this is known as a ``descriptor" and is denoted by $\vec{\sigma} = (\hat{\sigma_x}, \hat{\sigma_y},  \hat{\sigma_z})$. By evolving $\vec{\sigma}$ according to the unitary evolution of the system, the descriptor of a system at any given time gives us complete information about its state, when combined with knowledge of the initial statevector. 

The descriptor gives local and complete information about quantum states in the following sense: if we know the descriptor of two individual qubits, then we can recover the descriptor of their combined system, even if the two qubits are entangled. For example, if we know the descriptors of entangled qubits A and B individually, $\vec{\sigma_A}$ and $\vec{\sigma_B}$, we can recover their overall statevector $\ket{\psi}_{AB}$. Importantly, when quantum gates are applied to a system, it has been proven that the gates can only affect the descriptor of that quantum system; it cannot affect the descriptor of any quantum system on which the operation is not being directly applied \cite{deutsch2000information}. Hence, the existence of a local, complete description of quantum states using descriptors (i.e. evolved observables) shows explicitly that quantum theory satisfies the powerful constraints imposed by Einstein's principle of locality. 

\subsubsection{\textbf{Tracking information flow in the EPR thought experiment}}

\begin{figure}[htbp!]
\centerline{\includegraphics[width=0.5\textwidth]{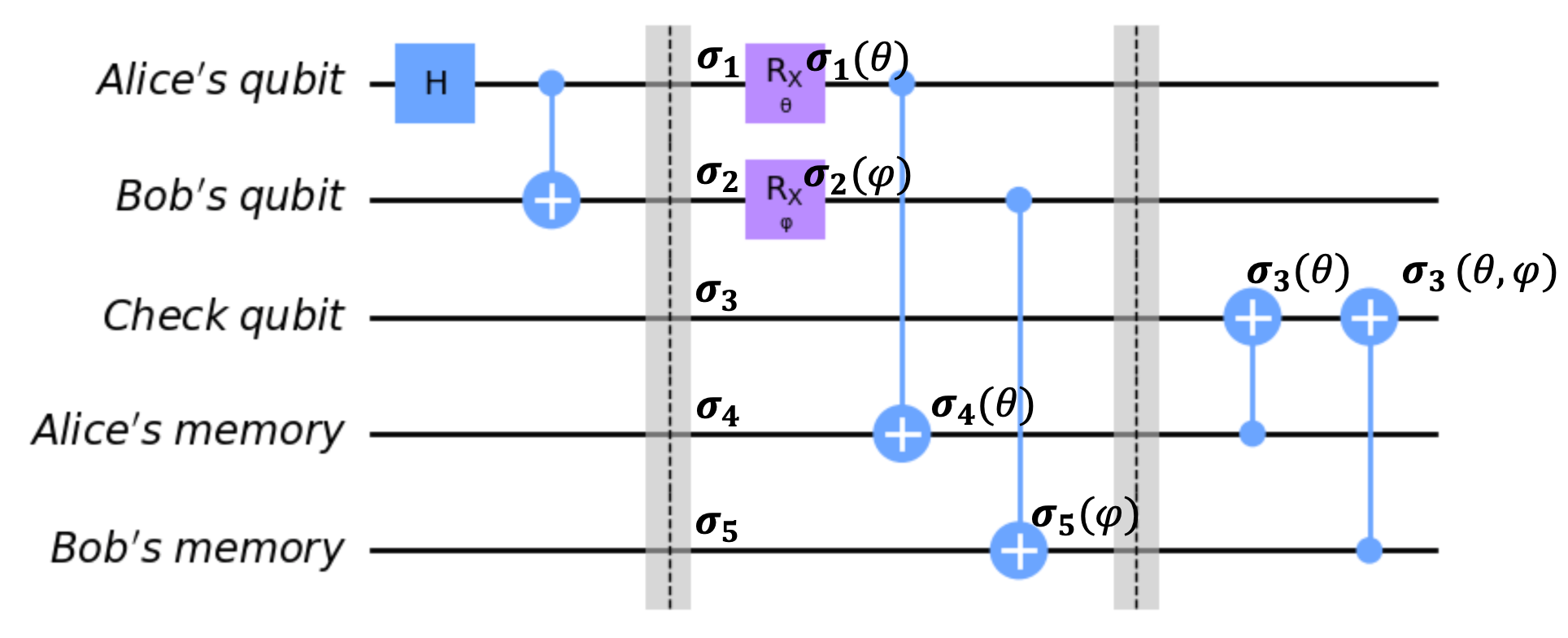}}
    \caption{Alice and Bob are treated as quantum systems, so all measurement operations are implemented via CNOT gates. Flow of the measurement basis information is shown using individual qubit descriptors.}
    \label{fig:info-flow}
\end{figure}

We can now apply this insight to the quantum circuit for the EPR thought experiment, to track the flow of information in the thought experiment, as shown in Figure \ref{fig:info-flow}. 

The figure shows how the descriptor of each qubit comes to depend on $\theta$ and $\phi$ from local interactions. We can see from the quantum circuit that there is a direct quantum operation between Alice's qubit and Alice's memory, which means that the descriptor of Alice's memory qubit can depend on the parameter $\theta$, which captures the basis she chose to measure her qubit in. Similarly, the observables describing Bob's memory qubit can depend on his basis-measurement parameter $\phi$. 

Next we have the two CNOT gates that implement the parity check with the check qubit. One of them acts directly between Alice's memory qubit and the check qubit, so the descriptor of the check qubit can now depend on $\theta$. The other acts directly between Bob's memory qubit and check qubit, so the check qubit's descriptor can also depend on $\phi$.

Therefore the final outcome of the check qubit, which reveals whether or not Alice and Bob's qubits are correlated, can depend on Alice and Bob's individual, independent choices of measurement basis. We can explain this dependence in a fully local way, tracking the flow of information about which bases Alice and Bob chose to measure their qubits in all the way through the circuit, step by step. Alice's choice of measurement basis had absolutely no distant influence on any aspect of Bob's qubit, and vice-versa; the correlations between their outcomes are only present when information about their basis choices has locally travelled from Alice and Bob's qubits to the check qubit. 

\subsection{Implications for special relativity and general relativity}

We can now take a step back and consider how far the behaviour of the entangled qubits, which we can test by running quantum circuits, is consistent with the key principles of special relativity and general relativity. 

An important principle for special relativity is no-signalling, and the related property of no faster-than-light communication. These conditions are based on the possibility of communication across distant parts of space, which has direct significance for laws surrounding causal influence. If changing one system could have an observable effect on a distant system, then that would enable communication at a distance, and associated paradoxes regarding causality. The quantum circuit analysis of an entangled pair of qubits, and the formulation in terms of descriptors, makes it abundantly clear that changes to individual systems of an entangled pair only affects that system alone, and therefore cannot cause issues with the principles of causality, no-signalling and no faster-than-light communication that are central to special relativity. 

Meanwhile the problem most pertinent for the research community relates to reconciling quantum theory with gravity, one of the greatest open problems of modern physics. This arises due to discrepancies between quantum theory and general relativity. The consistency of Einstein's principle of locality with quantum theory, via the local and complete description of entangled systems provided by descriptors, is an important property since it demonstrates a core principle that is obeyed by both theories. Having this core principle is significant for the broader programme of attempting to find a unifying, successor theory of quantum gravity. 

\subsection{Additional experiments with causality and locality}

There are many other fun and educational quantum computing activities that could be developed to investigate the laws surrounding causality. Some interesting directions include constructing simulations of indefinite causal orders, for example by implementing the ``quantum switch" protocol \cite{chiribella2013quantum, goswami2018indefinite}, or constructing a superposition of ``arrows of time" in opposing directions (which relates to our next activity, about entropy and the 2$^{\textrm{nd}}$ law of thermodynamics) \cite{rubino2021quantum}. Other protocols that can be used to investigate locality and the tracking of information flow include the quantum teleportation protocol, which is discussed in one of my instalments of the Quantum Paradoxes series \cite{violaris2023quantum}, and in several papers \cite{deutsch2000information, bedard2023teleportation}.

On the theme of Bell non-locality and its implications, I discuss a different experiment to Bell tests for deducing Bell non-locality, called Hardy's paradox, in another instalment of the Quantum Paradoxes series, framed as a game of Two Truths and a Lie \cite{violaris2023quantum}. There are also many interesting questions and results surrounding contextuality, which is a generalised form of Bell non-locality. Intriguing connections have been found between contextuality and the resources required for advantage for quantum computers, which could also form the basis of educational quantum computing activities for testing the resources required for computational advantage \cite{howard2014contextuality}. 

\section{Information erasure, entropy and the 2$^{\textrm{nd}}$ law}

In 1867, physicist James Clerk Maxwell wrote a letter to his friend Peter Guthrie Tait, unleashing a being that would haunt the foundations of thermodynamics to this day. In the letter, Maxwell explained a thought experiment involving a hypothetical clever little demon, that seems to be able to extract useful energy out of the random motion of the air. If true, this would violate the 2$^{\textrm{nd}}$ law of thermodynamics, by turning disorder into order, reducing the entropy of an isolated system. By merely using information about the particles, Maxwell's demon could in principle harness ambient heat to do useful tasks such as charging a mobile phone. This apparent violation of the 2$^{\textrm{nd}}$ law was resolved when scientist Charles Bennett realised the irreducible entropy cost of erasing information (Landauer's principle \cite{landauer1961irreversibility}) prevents Maxwell's Demon from turning heat into work in a cycle \cite{bennett1982thermodynamics}. 

The original thought experiment was entirely proposed and resolved classically. However, we know that classical models of reality have been superseded by quantum mechanics. When a system is modelled quantum mechanically, the meaning of work, heat and the 2$^{\textrm{nd}}$ law of thermodynamics become much more subtle. Understanding these properties is an area of study in the active and growing field of quantum thermodynamics \cite{vinjanampathy2016quantum}. A variety of approaches have been developed to quantify the 2$^{\textrm{nd}}$ law, work and heat for different physical situations, though there remains no consensus or universal approach for defining these properties on a quantum scale. Additionally, the thermodynamics of quantum systems has led to entirely new considerations, such as the thermodynamic implications of using the entropy stored in entanglement to do useful work \cite{rio2011thermodynamic}, and the role of other properties such as coherence \cite{lostaglio2015description}. 

A quantum Maxwell's Demon provides an ideal testing ground for analysing how work, heat and the 2$^{\textrm{nd}}$ law are manifested in quantum systems, and the physical meaning of and constraints on erasing quantum information. The constraints imposed on energy transfers by quantum thermodynamics has practical consequences for building quantum computers, for instance in considering the resources required for fault-tolerant quantum computing \cite{ben2013quantum} and to optimize trade-offs between energy dissipation, noise and memory resources required for algorithms \cite{meier2022thermodynamic}. 

I will start by explaining the classical Maxwell's Demon thought experiment; then Szilard's single-particle variation; and finally a quantum Maxwell's Demon presented as a quantum circuit. Then I will contextualise the findings in the field of quantum thermodynamics.  

\subsection{Classical Maxwell's Demon}

\begin{figure}[htbp!]
\centerline{\includegraphics[width=0.5\textwidth]{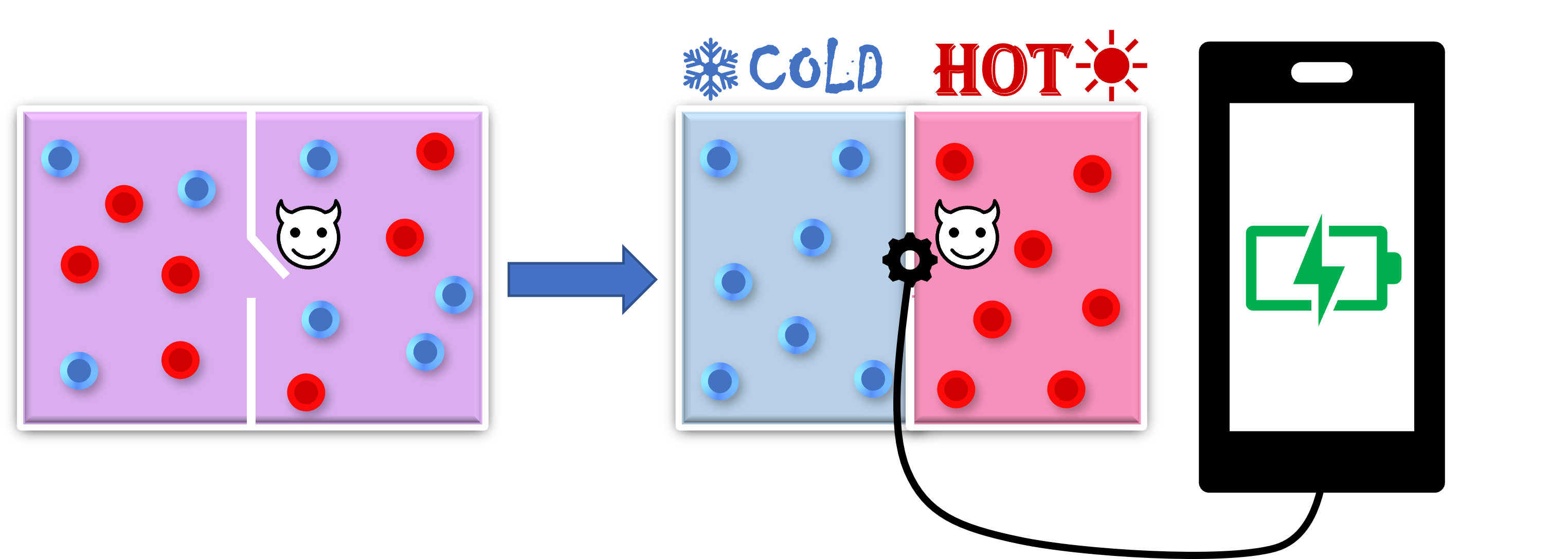}}
    \caption{Illustration of how demon can sort fast and slow particles into different sides of the box, leading to a temperature gradient that could be used to extract useful work, e.g. to charge a mobile phone.}
    \label{fig:demon-phone}
\end{figure}

In the classical Maxwell's Demon thought experiment, the demon has a box of particles randomly moving around. It inserts a partition in the box, which has a trap-door in the middle, which the demon can open and shut to block or permit particles to pass through (see Figure \ref{fig:demon-phone}). When the demon sees a fast-moving particle, it lets it through to the right side of the box, and when the demon sees a slow-moving particle, it lets it through to the left side. After a while, the right side will be filled with fast-moving particles, and the left side with slow-moving particles. The temperature of the right side is higher than the left, meaning the demon has created a temperature difference from a box originally at uniform temperature. This temperature difference can be used to power an engine and extract useful work. Hence, it seems that Maxwell's Demon could sort the particles in the air, and use them to generate electricity that can charge a mobile phone. This appears to violate the 2$^{\textrm{nd}}$ law of thermodynamics, which states that it is impossible to soley turn heat (random motion e.g. of the air) into work (useful energy e.g. that can charge a battery). 

\subsection{Resolution with Landauer's principle}

\begin{figure}[htbp!]
\centerline{\includegraphics[width=0.5\textwidth]{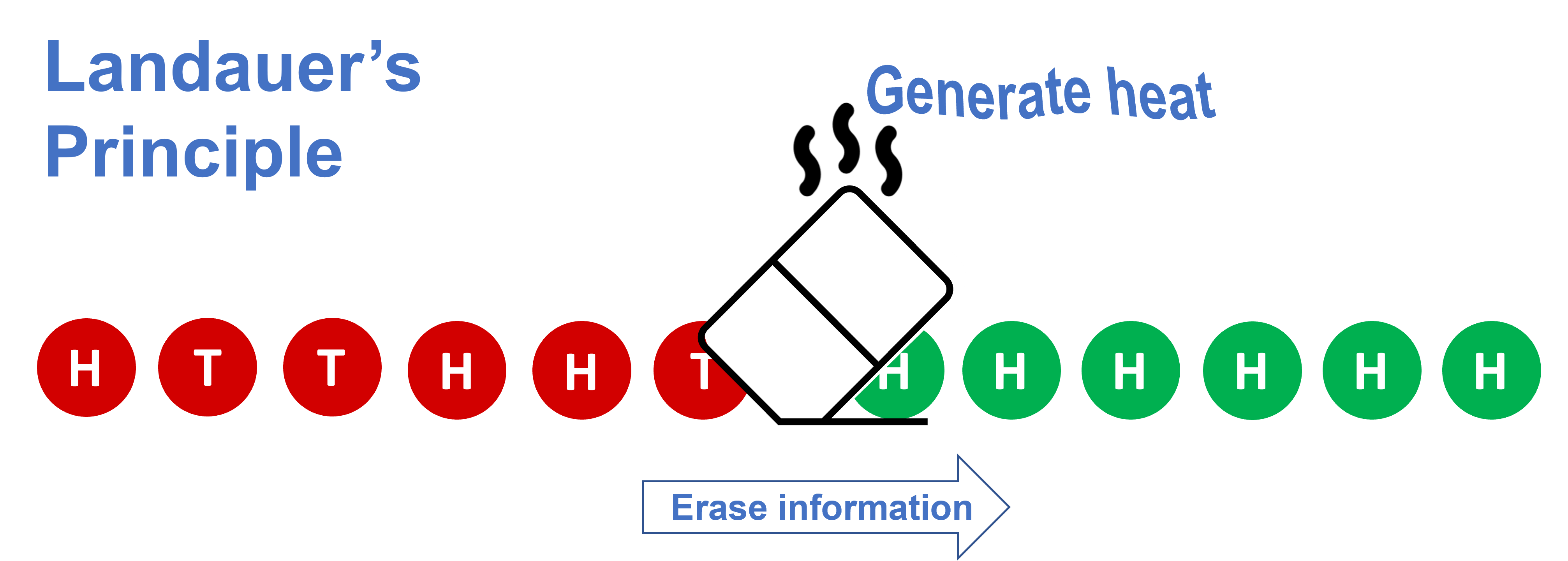}}
    \caption{A visualisation of Landauer's principle. Coins storing information by being in the Heads or Tails state can be ``reset" by the eraser to be in the blank Heads state. According to Landauer's principle, this type of information erasure has an irreducible entropy cost, to the environment, typically manifested as dissipation of heat.}
    \label{fig:landauer}
\end{figure}

First attempts to resolve the paradox, by Leo Szilard, focused on the demon's process of measuring the properties of the particles \cite{szilard1929entropieverminderung}. It was believed that there is a minimum work cost for the demon to retrieve information about the particle moving quickly or slowly, and the work needed for this step compensates the work gained from sorting the particles. However, in 1981 Charles Bennett showed that the measurement can actually be done with an arbitrarily small work cost \cite{bennett1982thermodynamics}. Instead, the solution came from Landauer's principle, which states that erasing information has a minimum work cost (see Figure \ref{fig:landauer}).

The demon needs to store the information it retrieves about the particles in its memory, which is finite. To extract work from the particles in a cycle, the demon's memory needs to be reset to its initial, blank state. According to Landauer's principle, this information erasure has a minimum work cost. Bennett pointed out that this work cost balances the work gained by sorting the particles, and therefore saves the 2$^{\textrm{nd}}$ law of thermodynamics. 

\subsection{Szilard's engine}

\begin{figure}[htbp!]
\centerline{\includegraphics[width=0.4\textwidth]{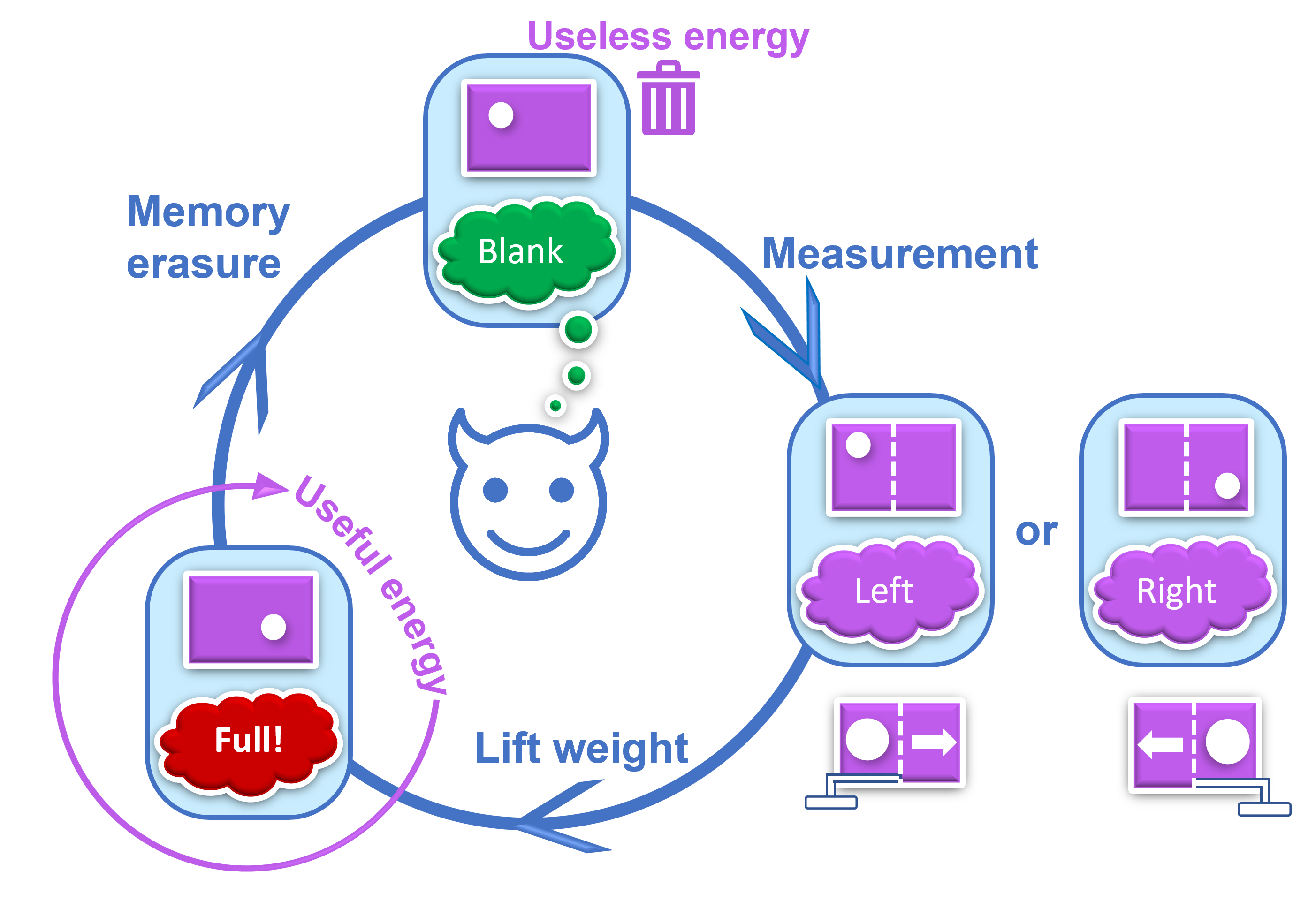}}
    \caption{Depiction of the Szilard's engine cycle, which is a single-particle realisation of Maxwell's Demon.}
    \label{fig:Szilard}
\end{figure}

To investigate Maxwell's Demon in a quantum setting, it is useful to first consider the classical single-particle version of the thought experiment. This is known as Szilard's engine, as it was proposed by Leo Szilard in 1928, and the cycle is depicted in Figure \ref{fig:Szilard} \cite{szilard1929entropieverminderung}.

Here we imagine there is just a single particle in the box, and the demon measures which side of the box the particle is on. Based on the measurement outcome, the demon inserts a partition, and a weight hanging from the same side of the box as the particle. As the particle expands to fill the full volume of the box again, it lifts the weight. During this process, the particle gains energy from exchanging with particles in its surroundings, so energy is conserved. The demon stored information about which side of the box the particle was on in its memory, in order to lift the weight and turn the random motion of the particle into useful energy. If we consider the demon to only have started with one bit of blank memory, then its memory needs to be reset to repeat the cycle. According to Landauer's principle, this has a minimum work cost. 

\subsection{Quantum Szilard's engine 
}
\begin{figure}[htbp!]
\centerline{\includegraphics[width=0.5\textwidth]{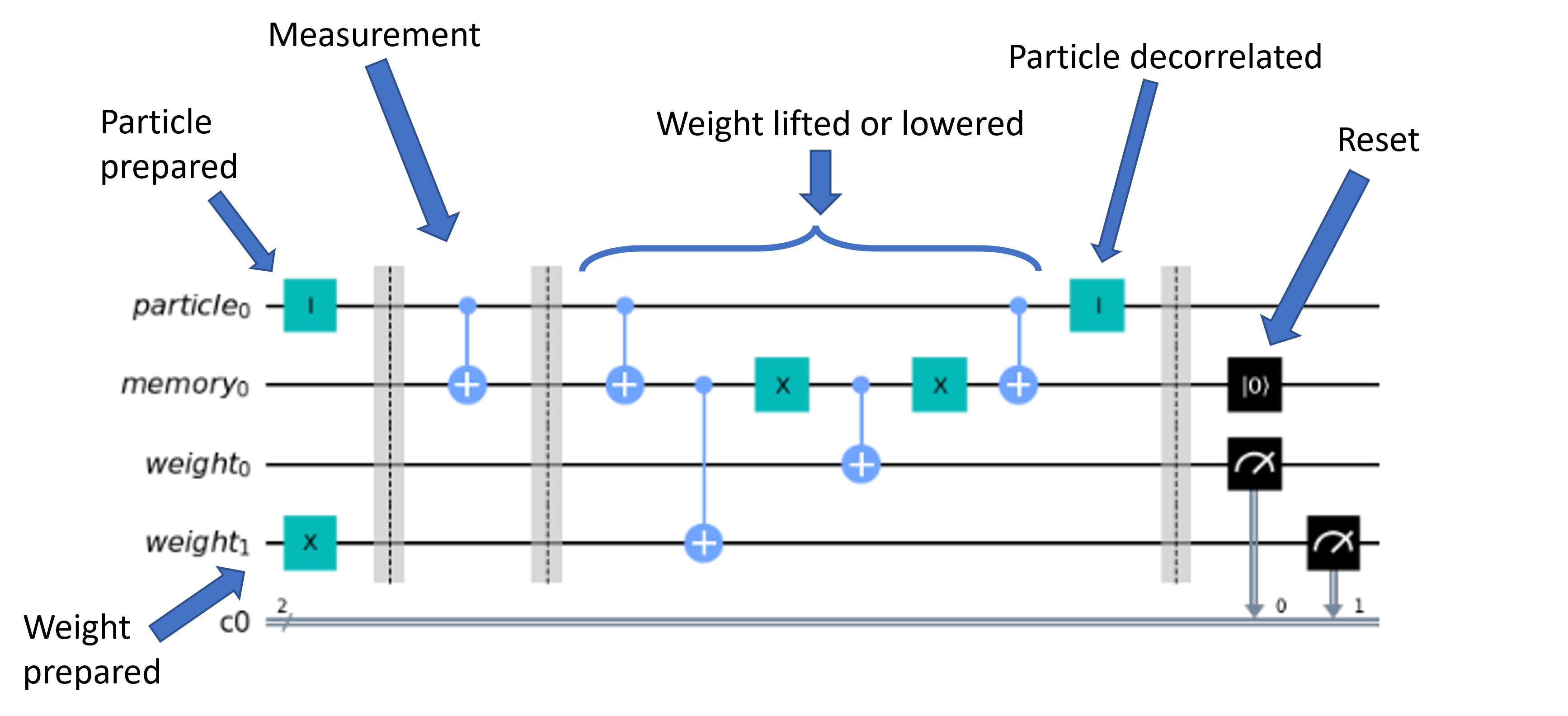}}
    \caption{The quantum circuit implementation of a Szilard's engine.}
    \label{fig:labelled-circuit}
\end{figure}

We can map each element of Szilard's engine to a quantum counterpart. We need four qubits in total for our quantum circuit implementation: a qubit for the particle in the box; a qubit for the demon's memory; and a pair of qubits for the weight that can be lifted or lowered. The full quantum circuit is depicted in Figure \ref{fig:labelled-circuit}, which we will now walk through step by step.

\subsubsection{Preparation of particle and memory}

The single particle randomly moving in the box can be modelled by a qubit in a maximally mixed state, which means its state is equivalent to having a 50$\%$ chance of being in the state $\ket{0}$ (i.e. left side of the box) and 50$\%$ chance of being in the state $\ket{1}$ (i.e. the right side of the box), with density matrix $\rho = \frac{1}{2}(\ket{0}\bra{0}+\ket{1}\bra{1})$. This could be prepared e.g. by flipping a coin to decide whether to prepare a $\ket{0}$ or $\ket{1}$ state. Note that this is distinct from the state $\frac{1}{\sqrt{2}}(\ket{0}+\ket{1})$, as the maximally mixed state has no coherence or superpostion between $\ket{0}$ and $\ket{1}$, it is just a random mixture of them.

For the Qiskit quantum circuit implementation, I prepared the particle in a maximally mixed state by adding an Identity gate to the circuit, and then using a depolarising noise model to add noise to the Identity gate, such that it becomes a quantum channel for mapping any state to a maximally mixed state. Note that in a practical physical implementation of Szilard's engine, rather than a simulation of the process, the particle would already begin in the maximally mixed state, and not need an additional preparation step. 

Meanwhile the initially blank memory of the Demon can be modelled by a qubit in a pure state, e.g. $\ket{0}$. We will see in a moment why it is key that for storing information that the qubit is in a pure state. Since qubits in the quantum circuit start in the $\ket{0}$ state by default, we do not add any additional gates to prepare the memory qubit. 

\subsubsection{Demon measures particle}

In the first step of the protocol, the demon measures the particle and sees whether it is on the left or right side of the box. As discussed in section \ref{epr-circuit}, we can model a measurement where the measuring apparatus is a quantum system using a CNOT gate, controlled on the system being measured and targeted on the measuring apparatus.

\subsubsection{Weight lifted or lowered}

In the classical Szilard's engine, the next step involves the demon using the information about which side of the box the particle is on to attach a weight on the same side, and then allowing the weight to be lifted by the pressure of the particle expanding into the box. For a qubit implementation of the weight, we can consider a system where the $\ket{1}$ state has high energy and the $\ket{0}$ state has low energy. A physical implementation of such a system could for instance be via the energy levels of an atom, with the $\ket{0}$ state being a lower energy level and $\ket{1}$ being an excited, higher energy level.

Now in order to see what happens when the weight is attached to the wrong side of the box (i.e. an error is made), we will also need to consider the weight being lowered rather than lifted, causing a loss in its ability to do useful work. Hence, we need our weight to have at least three possible energy states. This could be done using a qutrit (a three-level quantum system), however since we are working with a qubit quantum circuit, we can instead implement three energy levels by representing our weight using two qubits. Then, the state $\ket{00}$ is lowest energy, $\ket{01}$ and $\ket{10}$ are intermediate energies, and $\ket{11}$ is highest energy. We will start the cycle with the weight having intermediate energy level, i.e. the two-qubit state $\ket{01}$, so that the weight could be lifted or lowered during the cycle. Hence, we prepare the quantum weight by adding an X gate to one of the qubits. 

Now we need to add gates to implement the following logic: when the demon correctly attaches the weight to the side of the box with the particle, then the weight is lifted, so the weight qubits are mapped from $\ket{01}$ to $\ket{11}$. This happens when the particle qubit and memory qubit are in the same state. However, if the demon makes an error and attaches the weight to the wrong side of the box, then as the particle expands into the box the weight is lowered. In this case, the weight qubits are mapped from $\ket{01}$ to $\ket{00}$. This happens when the particle qubit and memory qubit are in different states, i.e. are anti-correlated.  The sequence of four doubly-controlled X-gates can be decomposed into the series of CNOTs and X-gates shown in the circuit. 

\subsubsection{Decorrelation of the particle and memory}

When expanding into the Szilard's engine box, the particle equilibriates with its environment and becomes decorrelated from the demon's memory. Hence it ends up in a maximally mixed state. We can implement this step in the circuit by again adding an Identity gate, to which we can add a maximally depolarising noise channel when running the circuit in Qiskit. 

\subsubsection{Erasing information}

Finally, we have the key step required to repeat the cycle and save the 2$^{\textrm{nd}}$ law of thermodynamics: the resetting of the memory state, with an irreducible entropy cost to the environment. In the quantum circuit, we can implement this by adding the non-unitary operation which resets a qubit from any state to a $\ket{0}$ state, which is our blank memory state. 

\subsubsection{Testing the significance of information erasure}

This quantum circuit implementation can now be used to test what happens with and without the memory erasure step, demonstrating the significance of the demon's memory qubit beginning in a pure state rather than a mixed state. If we include the reset operation, then the demon's memory ends in the $\ket{0}$ state, which is the state it was in originally. Running the cycle again will lead to exactly the same process of work extraction by lifting the weight, followed by work required to reset the demon's memory. 

If we skip the demon's memory reset step, then at the beginning of the next cycle, the demon's memory qubit is in a maximally mixed state. Then, after the measurement step, the memory qubit and particle qubit are not necessarily correlated: half the time they are correlated, in the same state, and half the time they are anti-correlated, in opposite states. If they happen to be correlated in a given run of the experiment, then it will work as before, and work will be extracted by lifting the weight. However, half the time they will be anti-correlated, and the weight will be lowered, meaning work is used up rather than extracted. Hence, on average, no work is extracted by repeated runs of the protocol. The demon's memory is useless when full, showing why the full Szilard engine cycle requires entropy to be spent in resetting the demon's memory. 

\subsubsection{Further activities and extensions}

One limitation to this model for the quantum Szilard engine is that, while it implements the right logic for the engine, it does not explicitly incorporate the thermodynamic constraint of energy conservation. The model could be augmented with additional qubits to track the energetic degrees of freedom of the particle, memory, and environment, showing how energy is conserved in the cycle. More broadly, by including the environment of the particle and the resource for resetting the memory explicitly in the circuit, one could create a fully unitary implementation of the quantum Szilard's engine. This would make manifest the resources required for the key steps to operate in a cycle, which could be particularly interesting as a pedagogical tool for learning about ``resource theories" as applied to thermodynamics and more generally \cite{lostaglio2016resource}. These theories formalise how different resources constrain the possibility of quantum transformations. Additionally, one could make a neater quantum circuit implementation of the cycle by making use of a qutrit for the weight system, on a platform where coding and implementing qutrits is supported.

There are a variety of exciting quantum thermodynamics directions that could be further explored with quantum circuit simulations. For example, implementing a quantum refrigeration cycle \cite{linden2010small}; a quantum heat engine \cite{quan2007quantum, ono2020analog}; and exploring various non-trivial additional bounds and extensions of Landauer's principle in the quantum setting \cite{van2022finite, Reeb_2014, esposito2010entropy}. Further applications to quantum technologies that could be explored include the possibility of using particular quantum logic to achieve cooling \cite{schulman2005physical} or exploring applications of autonomous quantum thermal machines \cite{lipka2023thermodynamic, guzman2023divincenzo}. 

\section{Time-travel paradoxes on quantum computers}

More than just a common trope in science fiction, time-loops that enable backwards-in-time travel (technically termed ``closed-timelike-curves", which we abbreviate to CTCs) are admitted by general relativity. However, this leads to various paradoxes, including both logical contradictions and violations of fundamental physics principles. 

Surprisingly, quantum computing provides a resolution to the famous ``Grandfather Paradox" of backwards time-travel, demonstrating how quantum CTCs can exist without logical contradictions \cite{deutsch1991quantum}. Furthermore, simulating CTCs using quantum circuits that we can run on current devices is an interactive way to see how they give advanced computational power beyond what is allowed by conventional quantum computers. This also serves as an interesting way to engage learners with the theory of computational complexity classes and their relation to physical phenomena.  

Finally the quantum circuit implementations elucidate open problems surrounding CTCs, whereby they appear to violate fundamental physical principles regarding locality and knowledge-creation \cite{deutsch1991quantum}. In addition to exploring the interesting solutions and problems raised by simulating CTCs on quantum computers, the activity in this section lays the groundwork for further quantum circuit implementations of spacetime thought experiments and phenomena. Simulations related to the foundations of spacetime could be a fruitful direction for near-term exploration using quantum computers, as demonstrated by recent experiments \cite{zhu2020generation, jafferis2022traversable}.

\subsection{\textbf{The Grandfather Paradox}}

The Grandfather Paradox is the most famous form of a time-travel paradox. A time traveler travels to the past and kills their own grandfather, before the time-traveler's parent has been born. This means the time traveler was never born in the first place, and therefore never went back in time to kill their grandfather, causing a contradiction.  A solution to the Grandfather Paradox in quantum mechanics was proposed by David Deutsch in 1991, using a consistency condition \cite{deutsch1991quantum}. This condition imposes that a system coming out of a time-machine must be identical to the one that went in. It holds whether the system in the time-loop is a qubit, or any other kind of quantum system. 

It is possible to do the trick of adding the consistency condition with classical systems going through a time-loop, not just quantum ones. The issue is that the classical systems only satisfy the consistency condition for very specific states, whereas in the quantum case, any state and interaction between future and past selves can satisfy the consistency condition. 

In terms of the Grandfather Paradox thought experiment: the classical consistency condition would mean that e.g. people that would kill their grandfather before their birth do not time-travel, or similarly people that time-travel are constrained by the dynamics of the universe not to be able to kill their grandfather. Both these options lead to physics in regions with time-loops enabled being vastly different to physics in regions without time-loops. While logically possible, this conflicts with our usual assumptions about the homogeneity of physics. 

By contrast, in the quantum case, people that wish to kill their grandfather before they are born can indeed travel back in time and do so. Logical contradictions are avoided because in this situation, the structure of branches in the quantum multiverse is such that the branch where the grandfather is killed is separate from the branch where the grandchild that kills him is born. Hence, imposing self-consistency in time-loops in a universe where everything obeys the dynamics of quantum theory does not restrict the choices of time-travellers, resolving any issues of having unusual physical laws in regions of spacetime with and without time-loops. 

As well as solving various problems, the consistency condition leads to a lot of powerful phenomena beyond standard quantum mechanics. For example, with access to a quantum time-loop, you can violate Heisenberg's uncertainty principle and the no-cloning theorem; store unbounded amounts of classical information in a single qubit; and break the security of quantum cryptographic protocols like BB84. 

\subsection{\textbf{A time-loop as a quantum circuit}}

\begin{figure}
    \centering
    \includegraphics[width=0.4\textwidth]{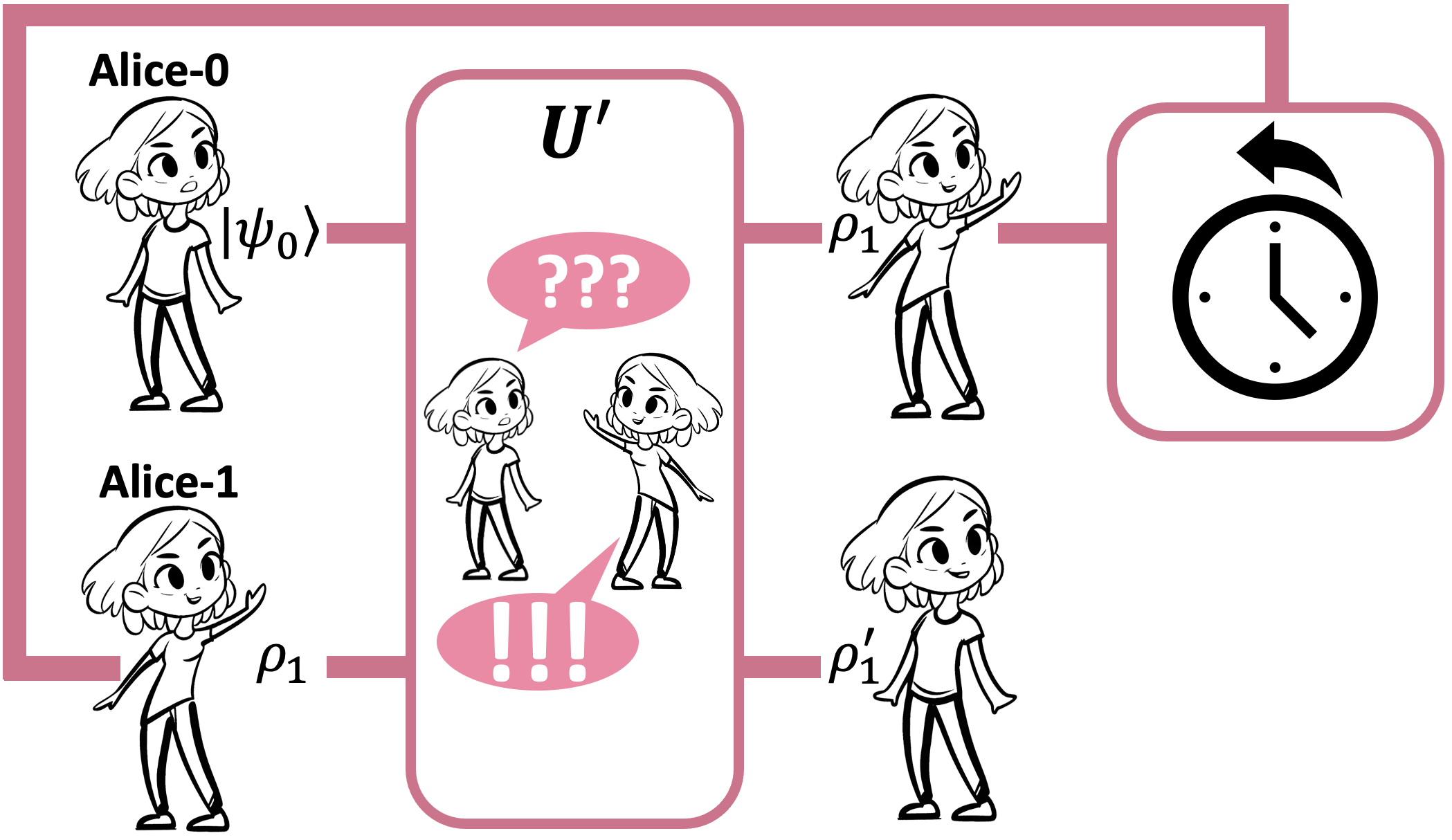}
    \caption{A visualisation of a time-loop, where Alice goes back in time to have a conversation with her past self.}
    \label{fig:Alice-timeloop}
\end{figure}

Consider the following thought experiment: a time-traveller goes into a time-machine which takes them back in time to meet their past self. Let's call the time-traveller Alice-1, and her past self Alice-0. Alice-1 steps into the time-machine, and goes an hour back in time. Stepping out of the time-machine, she finds her past self from 1 hour earlier, Alice-0. They chat for an hour, and then Alice-0 steps in to the time-machine. Alice-1 continues her life, while Alice-0 goes back in time to meet her own past self, so this time-loop continues (visualised in Figure \ref{fig:Alice-timeloop}). Alice-1 meeting and interacting with Alice-0 is allowed by the consistency condition, providing that Alice-1 comes out of the time-machine in the exact same state she was going into the time-machine. 

To map this onto a quantum circuit, we need to treat Alice as a quantum system. Quantum theory must apply to macroscopic observers if it is a universal theory; if you want to find out more about the validity of treating macroscopic observers as quantum systems and how to avoid inconsistencies, see my previous paper on quantum thought experiments as an educational tool, where I discuss the Schrödinger's cat and Wigner's friend thought experiments \cite{violaris2023physics}. We will assume that quantum theory is universal, and so the correct description of any system is given by its quantum state. 

In particular, the system may be in a quantum state that is pure (describable by a statevector $\ket{\psi}$, with density matrix $\rho_{\textrm{pure}} = \ket{\psi}\bra{\psi}$) or mixed (describable only by its density matrix $\rho_{\textrm{mixed}}$).  We will be using reduced density matrices $\rho$ to describe the states of quantum systems. Physically a reduced density matrix encapsulates all the information that is retrievable about an individual quantum system via measurements on that system alone. 

Let's say Alice-0 is in the state $\ket{\psi_{0}}$ (equivalently, $\rho_0 = \ket{\psi_{0}}\bra{\psi_{0}}$), when Alice-1 comes out of the time-machine to meet her. We will denote Alice-1's state by the density matrix $\rho_1$. Now Alice-0 and Alice-1 interact. We can model their interaction by a unitary gate $U'$. Then, Alice-0 enters the time-machine. To avoid paradoxes, Alice-0 must enter the time-machine in exactly the same state as Alice-1 came out of the time-machine. Hence, following Alice-0's interaction with Alice-1, Alice-0 must be in the state $\rho_1$. Meanwhile Alice-1 is in some arbitrary final state, which we can denote $\rho_1'$. The states are depicted on the diagram in Figure \ref{fig:Alice-timeloop}

\begin{figure}
    \centering
    \includegraphics[width=0.4\textwidth]{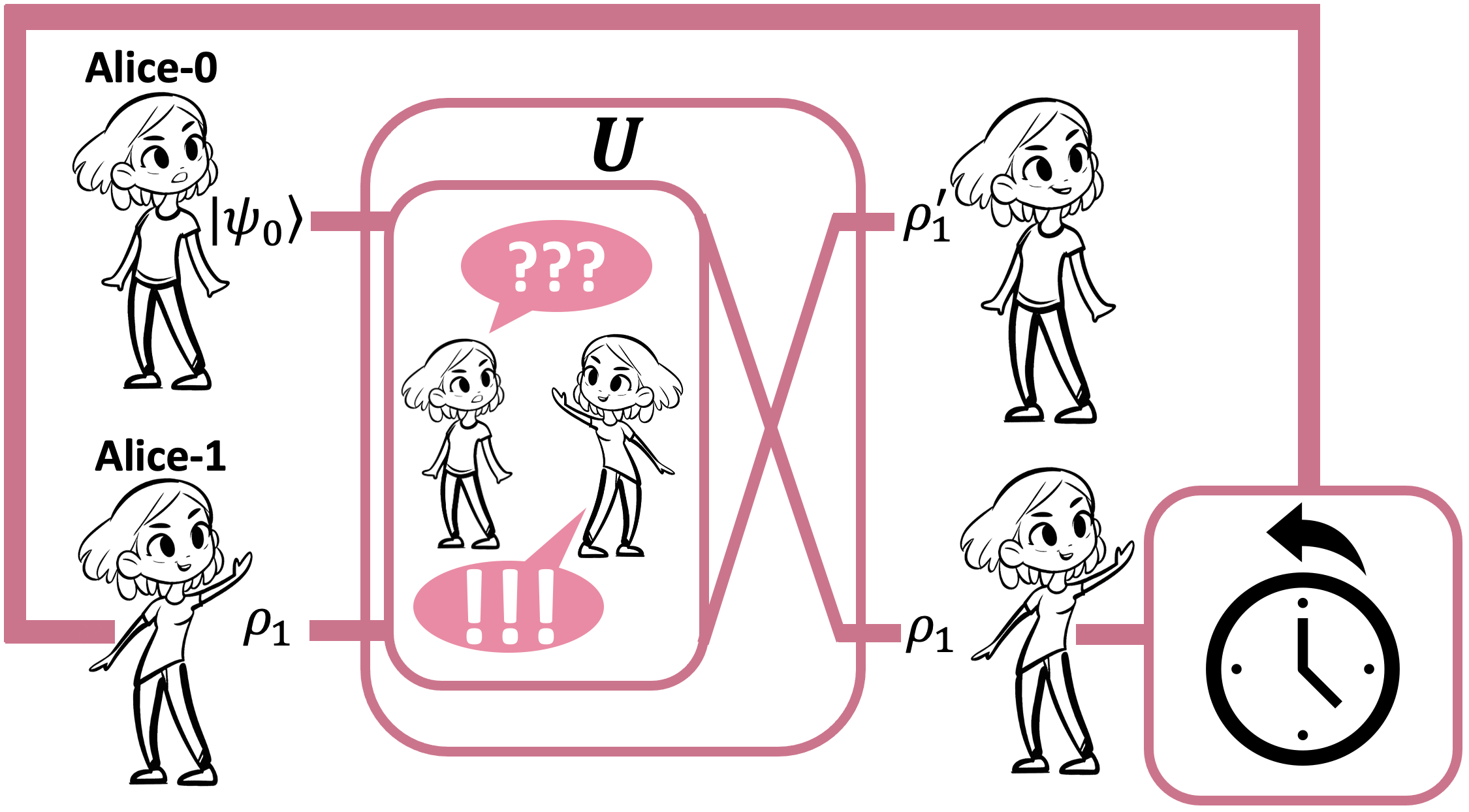}
    \caption{A visualisation of Alice's time-loop, with the SWAP gate absorbed into the interaction of Alice and her past self.}
    \label{fig:Alice-swap}
\end{figure}

It is helpful to rewrite this circuit in an equivalent form. Let's add a SWAP gate between Alice-0 and Alice-1 before the end of the circuit. This is a valid unitary gate, which swaps Alice-0 with Alice-1. Then we can rewrite the circuit using a unitary $U$, which is the same as $U'$ but with a SWAP gate at the end. Now the first qubit begins in $\rho_0$ (Alice-0's initial state) and comes out in $\rho_{1}'$ (Alice-1's final state) after the unitary $U$, and the second qubit comes out of the time machine in the state $\rho_{1}$ (Alice-1's initial state), then goes back in to the time-machine in the state $\rho_{1}$ (Alice-0's final state), after the unitary $U$. This is visualised in Figure \ref{fig:Alice-swap}.

So in our rewritten circuit, the consistency condition for avoiding a version of the Grandfather Paradox is that the qubit that goes into the time-loop must be in the same state as the qubit that comes out of the time-loop. Mathematically, the reduced state of the second qubit in our modified circuit must be the same before and after the unitary gate, i.e., $\rho_{1} = \text{tr}_1 \left( U \left( |\psi_{0}\rangle \langle \psi_{0} | \otimes \rho_{1} \right) U^\dagger \right)$. For what follows, we will refer to the first qubit as the ``system qubit", and the second qubit as the ``time-loop qubit". 

Now, famously, standard quantum mechanics is linear. The consistency condition introduces non-linearity, because the consistent state imposed on the qubit in the time-loop depends on the state of the qubit it interacts with. This non-linearity is the source of the power of computing with access to time-loops, over and above computing that is possible with standard quantum computers.

\subsection{\textbf{Coding with Closed-Timelike-Curves}}

Unfortunately, we do not currently have access to genuine time-loops, and it is unknown whether they can even really exist. Hence, to interactively explore their power, we will use simulations where we artificially insert the non-linearity. 

\subsubsection{\textbf{Distinguishing $\ket{0}$ and $\ket{-}$ with one measurement}}

\begin{figure}
    \centering
    \includegraphics[width=0.3\textwidth]{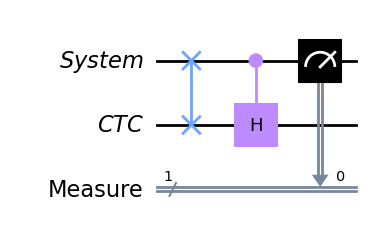}
    \caption{The quantum circuit for distinguishing $\ket{0}$ and $\ket{-}$, when the user input is ``0".}
    \label{fig:distinguish}
\end{figure}

For our first trick with CTCs, we see how they let us distinguish the quantum states $\ket{0}$ and $\ket{-}$ with a single measurement. This protocol was introduced in \cite{brun2009localized}. One way of coding a simulation of this, as I implement in my Qiskit video and code tutorial \cite{violaris2023quantum}, is to ask the user for an input of ``0" to prepare the $\ket{0}$ state and ``-" to prepare the $\ket{-}$ state. If the user inputs ``0", then my code leaves the system qubit as $\ket{-}$ and the time-loop qubit as $\ket{0}$, which is the state needed to make the time-loop qubit's input and output states the same according to the consistency condition. The code is:
\begin{lstlisting}
from qiskit import QuantumCircuit, QuantumRegister, ClassicalRegister
user_input = input("Enter '0' to prepare |0> or '-' to prepare |->")
sys = QuantumRegister(1, "System")
ctc = QuantumRegister(1, "CTC")
cr = ClassicalRegister(1, "Measure")
qc = QuantumCircuit(sys, ctc, cr)
if user_input == "-":
    qc.h(sys)
    qc.z(sys)
    qc.barrier()
    qc.x(ctc)
    qc.barrier()
qc.swap(sys, ctc)
qc.ch(sys, ctc)
qc.measure(sys, cr)

\end{lstlisting}
If the user inputs ``-", then my code applies an H and Z gate to the system qubit to prepare it in the $\ket{-}$ state, and an X gate to the time-loop qubit to prepare it in the $\ket{1}$ state, which is the state needed to satisfy the consistency condition when the input is $\ket{-}$. I have artificially introduced the non-linearity into the quantum circuits by making the preparation of the time-loop qubit depend on the user's input state. In a real time-loop, these would automatically be self-consistent, which gives it the extra computational power.  Then, if you enter ``0", the output is always 0, and if you enter ``-", the output is always 1: so the time-loop allows us to perfectly distinguish the ``0" state from the ``-" state with a single measurement. The quantum circuit for the ``0" input is shown in Figure \ref{fig:distinguish}.

\subsubsection{\textbf{Distinguishing states for BB84 protocol}}

\begin{figure}
    \centering
    \includegraphics[width=0.5\textwidth]{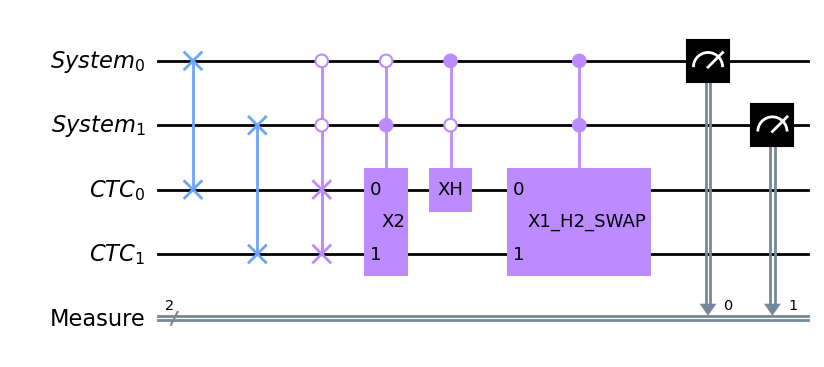}
    \caption{The quantum circuit for distinguishing $\ket{0}$ and $\ket{-}$, when the user input is ``00".}
    \label{fig:bb84}
\end{figure}

Using the same idea, we can create a quantum circuit that can distinguish between the four states $\ket{0}$, $\ket{1}$, $\ket{+}$ and $\ket{-}$ in a single shot. This means that someone with access to a time-loop could cheat the BB84 protocol, which is meant to provide a fool-proof way of protecting information from eavesdroppers using quantum cryptography. The protocol relies on the ability to check if someone intercepted a message, but with a time-loop, the eavesdropper could intercept the message, measure its state, and then prepare an exact copy of the state, covering their tracks. By coding this in a similar way to the previous circuit, we get the following behaviour: when we input 0, 1, + and -, the outputs are 00, 10, 01 and 11 respectively. Figure \ref{fig:bb84} shows the quantum circuit when the user inputs 00. So, a single measurement will distinguish any of the four possible input states. 

\subsection{\textbf{The strange powers given by CTCs}}

From an information-theory perspective, we would achieve something extremely surprising if we could really implement this circuit. We would gain two bits of classical information by measuring a single input qubit, even though a qubit can normally only store 1 bit of information. Even worse, we could construct a similar circuit to distinguish between any number of non-orthogonal states. This means we could store an unbounded amount of classical information in a single qubit, and also violate the no-cloning theorem \cite{ahn2013quantum}.

Access to time-loops make us much more powerful than if we just had access to a universal quantum computer \cite{aaronson2009closed}. The power of computers is defined by complexity classes. Aaronson and Watrous showed that the complexity class for a computer with access to time-loops is PSPACE: all problems solvable by a conventional computer using a polynomial amount of memory (though these problems may still require exponential time to solve). PSPACE is a huge space, much bigger than the complexity class of quantum computers without time-loops. Strangely, given access to time-loops, the complexity classes of classical and quantum computers become the same.

\subsection{\textbf{Paradoxes remain}}

Some problems remain in making time-loops fully consistent with the laws of physics. One is the ``Knowledge Paradox." Continuing our previous time-travel example, imagine that Alice-1 explains a mathematical result to her past self, Alice-0. Armed with this knowledge, Alice-0 then steps into the time-machine and explains the mathematical result to her past self. There is no logical inconsistency here, but where did the knowledge about the mathematical result come from? If Alice was taught about time-loops from her past self, then there is no mechanism by which that knowledge could have been created.

Another problem is that observables of a qubit going into a time-loop could in principle be different from those coming out of the time-loop, even though the density matrix is the same, violating Einstein's locality principle \cite{kuypers2022unorthodox}.

\subsection{\textbf{Extensions to exploring time-loops}}

Further quantum circuit simulations of time-loops could demonstrate violating the no-cloning theorem, and storing large amounts of classical information in a single qubit. Another direction could be to explore activities based on time-loops that are relevant to recent research in quantum technologies, e.g. \cite{arvidsson2022quantum}, which explores a closed-timelike-curve inspired approach to quantum metrology. 

\section{Summary and conclusion}

We have explored how to use interactive quantum computing activities to explore foundational topics in physics, such as causality, entropy, and time-travel. The interplay between quantum computing and other disciplines of fundamental physics elucidates foundational principles of physics found in special relativity, general relativity and thermodynamics. Simultaneously, it helps teach fundamental aspects of the principles underlying quantum computing, and the constraints surrounding it. The activities are based on simple quantum circuit implementations that learners can use to practice and play with current quantum computers. For further details of code implementations of the activities, I refer the reader to the code tutorials linked in the associated Quantum Paradox video descriptions \cite{violaris2023physics}.

The quantum computing activities outlined in this paper, and further developments of quantum circuit simulations on these themes, could catalyse new insights and creations from learners. They provide hands-on avenues for learners to play with today's quantum devices; interactively explore interdisciplinary physical principles; and engage with topics on the edge of current research on foundations of physics. 

\section*{Acknowledgment}

MV thanks everyone in the community team at IBM Quantum who was involved in helping produce the Quantum Paradoxes content series, and all her academic colleagues in Oxford and Bristol who engaged in discussions about the quantum circuit implementation of Szilard's engine.

\bibliographystyle{IEEEtran}
\bibliography{references}  

\end{document}